\documentclass[12pt]{article}

\begin{document}

\title{A New Approximation to the Normal Distribution Quantile Function}

\author{Paul M. Voutier}
\date{}

\maketitle

\begin{abstract}
We present a new approximation to the normal distribution quantile function. It
has a similar form to the approximation of Beasley and Springer \cite{BS},
providing a maximum absolute error of less than $2.5 \cdot 10^{-5}$. This is less
accurate than \cite{BS}, but still sufficient for many applications. However it
is faster than \cite{BS}. This is its primary benefit, which can be crucial to
many applications, including in financial markets.
\end{abstract}

\section{Introduction}

The use of the inverse of the CDF for a probability distribution, also known as
the quantile function, is widespread in statistical modelling (see, for example,
\cite{G, J}).

During recent work, the need arose for a fast and reasonably accurate approximation
to the normal distribution quantile function, $N^{-1}(x)$. Accuracy similar to
the approximation in Equation~26.2.23 of \cite{AS} was sufficient (max absolute
error less than $4.5 \cdot 10^{-4}$). But speed was crucial.

The approximation of Beasley and Springer \cite{BS}, along with related
approximations such as Acklam's \cite{A}, provides improvements in terms of
both accuracy and speed.

Both the Acklam and the Beasley-Springer approximations are based on the same ideas:\\
(1) consider narrow tails separately from a wide central area\\
(2) use a rational function of $x$ to approximate $N^{-1}(x)$ in this wide central
area (avoiding expensive operations like $\log$ and sqrt)\\
(3) take advantage of the fact that $N^{-1}(x-1/2)$ is an odd function.

The second and third ideas suggest that for the central region, we consider rational
approximations of the form
$$
(x-1/2)F((x-1/2)^{2}),
$$
where $F$ is a rational function. The approximations of Acklam, Beasley-Springer,
and others for the central region are of this form.

The Beasley-Springer approximation for the central region is sometimes called a
$(3,4)$ scheme, since the numerator of $F$ is cubic in $(x-1/2)^{2}$ and the
denominator of $F$ is of degree $4$ in $(x-1/2)^{2}$. Similarly, the Acklam
approximation is called a $(5,5)$ scheme.

\section{New Approximations}

For increased speed, here we consider a $(2,2)$ scheme for the central region and
a $(3,2)$ scheme for the tails.

We chose the boundaries between the central region and the tails to be at $0.0465$
and $0.9535$, since with the above schemes and boundaries the maximum absolute error
in both regions was nearly the same and both slightly less than $2.5 \cdot 10^{-5}$.

\subsection{Central Region}

\subsubsection{$0.0465 \leq p \leq 0.9535$}

Put $q=p-0.5$ and let $r=q^{2}$. For $0.0465 \leq p \leq 0.9535$, define
\begin{displaymath}
f_{central}(p) =
q \frac{\displaystyle a_{2}r^{2}+a_{1}r+a_{0}}
{\displaystyle r^{2}+b_{1}r+b_{0}}
= q \left( a_{2} + \frac{\displaystyle a_{1}'r+a_{0}'}
{\displaystyle r^{2}+b_{1}r+b_{0}} \right)
\end{displaymath}
where
\begin{eqnarray*}
a_{0}  & = &  0.389422403767615, \\
a_{1}  & = & -1.699385796345221, \\
a_{2}  & = &  1.246899760652504, \\
a_{0}' & = &  0.195740115269792, \\
a_{1}' & = & -0.652871358365296, \\
b_{0}  & = &  0.155331081623168, \\
b_{1}  & = & -0.839293158122257.
\end{eqnarray*}

The benefit of the second expression is that we save one multiplication by using
it. Similarly, normalising the denominator so that the leading coefficient is $1$,
rather than the constant coefficient as some authors do, also saves another
multiplication.

There are 12 points of maximum error (also known as {\it alternating points})
in the interval $[0.0465, 0.9535]$:
\begin{center}
\begin{tabular}{ || l || l || }
\hline
$(p, err_{abs})$                     & $(p, err_{abs})$                     \\ \hline
$(0.046500, 2.494327 \cdot 10^{-5})$ & $(0.592289, 2.494326 \cdot 10^{-5})$ \\ \hline
$(0.054264, 2.494331 \cdot 10^{-5})$ & $(0.752182, 2.494327 \cdot 10^{-5})$ \\ \hline
$(0.081621, 2.494328 \cdot 10^{-5})$ & $(0.859308, 2.494323 \cdot 10^{-5})$ \\ \hline
$(0.140694, 2.494323 \cdot 10^{-5})$ & $(0.918381, 2.494328 \cdot 10^{-5})$ \\ \hline
$(0.247820, 2.494327 \cdot 10^{-5})$ & $(0.945738, 2.494331 \cdot 10^{-5})$ \\ \hline
$(0.407712, 2.494326 \cdot 10^{-5})$ & $(0.945350, 2.494327 \cdot 10^{-5})$ \\ \hline
\end{tabular}
\end{center}

From the theorems of Chebyshev and de la Vall\'{e}e Poussin (see \cite[Section~5.5]{C}), it
follows that $f_{central}(p)$ is essentially the best possible rational approximation
of $(2,2)$ scheme.

For comparison, the maximum absolute error of the ``central'' approximation in
\cite{BS} is under $1.85 \cdot 10^{-9}$.

This approximation was found using the minimax function within the numapprox
package of Maple:
\begin{verbatim}
Digits:=60:with(numapprox):
uBnd:=0.4535^2:
minimax(x->inverseCDFCentralRatApprox(x),0..uBnd,[2,2],x->sqrt(x));
\end{verbatim}
where\\
inverseCDFCentralRatApprox(x) is the function $N^{-1}(\sqrt{x}+1/2)/\sqrt{x}$,\\
uBnd is the range we want the approximation over,\\
$[2,2]$ specifies that we want the degree of both the numerator and the denominator
to be $2$, and\\
$\sqrt{x}$ is the weight function we use, since we want to get the best approximation
to $N^{-1}(\sqrt{x}+1/2)$ rather than $N^{-1}(\sqrt{x}+1/2)/\sqrt{x}$.

We tried other values of uBnd near $0.4535$, but the smallest maximum absolute error
was found with this particular value.

\subsubsection{$0.025 \leq p \leq 0.975$}

The use of an even wider central region may be preferred, as this can provide
further performance gains by reducing the expensive log and sqrt operations
required for the tails.

We give one such example here (found as above using Maple, but with uBnd=0.475).

Put $q=p-0.5$ and let $r=q^{2}$. For $0.025 \leq p \leq 0.975$, define
\begin{displaymath}
f_{central}(p)
= q \left( a_{2} + \frac{\displaystyle a_{1}r+a_{0}}
{\displaystyle r^{2}+b_{1}r+b_{0}} \right)
\end{displaymath}
where
\begin{eqnarray*}
a_{0}  & = &  0.151015505647689, \\
a_{1}  & = & -.5303572634357367, \\
a_{2}  & = &  1.365020122861334, \\
b_{0}  & = &  0.132089632343748, \\
b_{1}  & = & -.7607324991323768.
\end{eqnarray*}

The maximum absolute error for this approximation is less than $1.16 \cdot 10^{-4}$
which occurs near $p=0.9692$. While this error is much larger than the error in the
previous section, it is still well smaller than the maximum error for the
Abramowitz-Stegun approximation ($4.5 \cdot 10^{-4}$).

\subsection{Tails}

\subsubsection{$e^{-37^2/2} < p < 0.0465$}

For $5.3\ldots \cdot 10^{-298} = e^{-37^2/2} < p < 0.0465$, put $r=\sqrt{\log(1/p^{2})}$
and define
$$
f_{tail}(p) = \frac{c_{3}r^{3} + c_{2}r^{2} + c_{1}r + c_{0}}{r^{2} + d_{1}r + d_{0}}
= c_{3}r + c_{2}' + \frac{c_{1}'r + c_{0}'}{r^{2} + d_{1}r + d_{0}}.
$$
where
\begin{eqnarray*}
c_{0}  & = & 16.896201479841517652, \\
c_{1}  & = & -2.793522347562718412, \\
c_{2}  & = & -8.731478129786263127, \\
c_{3}  & = & -1.000182518730158122, \\
c_{0}' & = & 16.682320830719986527, \\
c_{1}' & = &  4.120411523939115059, \\
c_{2}' & = &  0.029814187308200211, \\
d_{0}  & = &  7.173787663925508066, \\
d_{1}  & = &  8.759693508958633869.
\end{eqnarray*}

As with the ``central'' approximation, this approximation was also found using the
minimax function within the numapprox package of Maple:
\begin{verbatim}
Digits:=60:with(numapprox):
v:=0.0465:
uBnd:=0.4535^2:
minimax(y->inverseCDF(exp(-y*y/2)), sqrt(log(1/v^2))..37, [3,2]);
\end{verbatim}
Note that since we are approximating $N^{-1}(x)$ itself here, we do not include
a weight function in the arguments of the minimax function and so the default
weight function $1$ is used.

The maximum absolute error in this case is less than $2.458 \cdot 10^{-5}$.

\subsubsection{$0.9535 < p < 1-e^{-37^2/2}$}

Due to the symmetry of $N^{-1}(p)$ about $p=1/2$, we approximate $N^{-1}(p)$
by $-f_{tail}(1-p)$ (note that here $r=\sqrt{\log(1/(1-p)^{2})}$).

\section{Abramowitz and Stegun Approximations}

Having found the above new approximations, we turned our attention to the approximations
in Equations~26.2.22 and 26.2.23 of \cite{AS}. As those authors note, these
approximations are from \cite{H}. In particular, Sheets 67 and 68 on pages~191--192
of \cite{H}.

If we restrict our attention to ranges like $e^{-37^2/2}< p < 1-e^{-37^2/2}$
(this includes almost the entire IEEE-754 range of representable real numbers),
then we can improve on the approximations of Abramowitz and Stegun.

For example, in this range, we can replace Equation~26.2.23 of \cite{AS} with
$$
x_{p}=t-
\frac{c_{2}t^{2} + c_{1}t + c_{0}}{d_{3}t^{3} + d_{2}t^{2} + d_{1}t + 1}
+\epsilon(p),
$$
where $|\epsilon(p)|<8 \cdot 10^{-5}$ and

\begin{eqnarray*}
c_{0} & = &  2.653962002601684482, \\
c_{1} & = &  1.561533700212080345, \\
c_{2} & = &  0.061146735765196993, \\
d_{1} & = &  1.904875182836498708, \\
d_{2} & = &  0.454055536444233510, \\
d_{3} & = &  0.009547745327068945.
\end{eqnarray*}

This is over five times more accurate than the approximation in \cite{AS}.
However, as one increases the range even closer to $0$ and $1$, the max absolute
increases until we obtain Equation~26.2.23 of \cite{AS}. The near-best possible
nature of Equation~26.2.23 is illustrated by the graph in Sheet~68 of \cite{H}
showing that Chebyshev's theorem nearly holds for this approximation.

Note also that this approximation shows the justification for the use of
$\sqrt{\log(1/p^{2})}$ in these tail approximations. As $p \rightarrow 0$,
$N^{-1}(p)$ approaches $-\sqrt{\log(1/p^{2})}$ plus a quantity that approaches $0$
as $p$ does.

\section{Performance}

Using Java (JDK $1.6.0\_17$), we coded the following approximations in order to compare
their performance.

\noindent
$\bullet$ the Abramowitz-Stegun approximation (AS in the table below)\\
$\bullet$ the Beasley-Springer approximation (BS in the table below)\\
$\bullet$ the approximation from Section~2 using the central region approximation in Section~2.1.1
(Rat22A in the table below)\\
$\bullet$ the approximation from Section~2 using the central region approximation in Section~2.1.2
(Rat22B in the table below).

In each case, we calculated the approximation 200,000 times for each $p$ from
$0.001$ to $0.999$ with $0.001$ as our step size. These calculations were done
on a Dell Inspiron 1525, running Windows Vista and using an Intel Core 2 Duo
T5800 2.00 GHz CPU. The times in milliseconds for each approximation are given
in the table below.

\begin{center}
\begin{tabular}{ | l | l | }
\hline
method & time(ms) \\ \hline
AS     & 25,210   \\ \hline
BS     & 10,212   \\ \hline
Rat22A & 8052     \\ \hline
Rat22B & 6649     \\ \hline
\end{tabular}
\end{center}

As one would expect, the new approximations given here are faster than the currently
known ones. The comparison between Rat22A and Rat22B is also interesting, as it shows
the impact of the calculation of the log and sqrt operations. Although these operations
only need to be performed for a small subset of all values of $p$, reducing the number
of these operations by just under 50\% reduced the CPU time required by nearly 20\%.

\bibliographystyle{amsplain}

\noindent
Paul Voutier\\
London, UK\\
\texttt{paul.voutier@gmail.com}
\end{document}